\documentclass[12pt]{iopart}
\usepackage{iopams,cite}
\usepackage{graphicx}
\usepackage{amsfonts}
\usepackage{mathrsfs}
\usepackage{epstopdf}
\usepackage{bm}

\newcommand{\ii}{\;\! \mbox{i} \;\!}

\newcommand{\mb}[1]{{\rm #1}}

\begin{document}

\title[Resonant tunneling of Bose-Einstein condensates in optical lattices]{Resonant tunneling of Bose-Einstein condensates in optical lattices}
\author{Alessandro Zenesini$^{\dagger}$, Carlo Sias$^{*}$, Hans Lignier$^{*}$, Yeshpal Singh, Donatella Ciampini$^{\dagger}$, Oliver Morsch$^{*}$, Riccardo Mannella$^{\dagger}$, Ennio Arimondo$^{*,\dagger}$}
\address{Dipartimento di Fisica Enrico Fermi, Universit\`{a} degli Studi di Pisa, Largo Pontecorvo 3, I-56127 Pisa}
\address{$^{*}$CNR-INFM, Largo Pontecorvo 3, I-56127 Pisa}
\address{$^{\dagger}$CNISM Unit\'a di Pisa, Largo Pontecorvo 3, I-56127 Pisa}
\author{Andrea Tomadin}
\address{Scuola Normale Superiore, Piazza dei Cavalieri 7, I-56126 Pisa}
\author{Sandro Wimberger}
\address{Institut f\"ur Theoretische Physik, Universit\"at Heidelberg,
Philosophenweg 19, D-69120 Heidelberg}

\begin{abstract}
In this article, we present theoretical as well as experimental
results on resonantly enhanced tunneling of Bose-Einstein
condensates in optical lattices both in the linear case and for
small nonlinearities. Our results demonstrate the usefulness of
condensates in optical lattices for simulating Hamiltonians
originally used for describing solid state phenomena.
\end{abstract}


\section{Introduction}
\label{intro}
In the last decade, the experimental techniques used in atom and
quantum optics have made it possible to control the external and
internal degrees of freedoms of ultracold atoms with a very high
degree of precision. Thus, ultracold bosons or fermions loaded
into optical lattices are optimal realizations of lattice
models proposed and studied in the context of solid-state physics.
Bose-Einstein condensates, for instance, have been used to
simulate phenomena such as Bloch oscillations in tilted periodic
potentials \cite{Peik1996,Raizen1997,Morsch2001,Cristiani2002,RMFOMI2004,MO2006}
and to study quantum phase transitions driven by atom-atom
interactions \cite{BDZ2007}.

Up to now most of the quantum transport phenomena investigated
with Bose-Einstein condensates within periodic optical lattices
focused on the atomic motion in the ground state band of the
periodic lattice. Only a few experiments examined the quantum
transport associated with interband transitions ``vertical'' in
the energy space. Interband transitions were induced by additional
electromagnetic  fields, as in the case of the spectroscopy of
Wannier-Stark levels \cite{R1996}, or by quantum tunneling between
the bands. Tunneling between otherwise uncoupled energy bands
occurs  when the bands are coupled by an additional force, which
can be a static Stark force (tilting the otherwise periodic
lattice) \cite{MO2006}, or also by strong atom-atom interactions
as observed for fermions in \cite{KMSGE2005} and discussed for
bosons in \cite{lee2007}. The quantum tunneling between the ground
and the first excited band is particularly pronounced in the
presence of degeneracies of the single-well energy levels within
the optical lattice leading to resonantly enhanced tunneling
(RET). RET is a quantum effect in which the probability for
tunneling of a particle between two potential wells is increased
when the energies of the initial and final states of the process
coincide. Owing to the fundamental nature of this effect and the
practical interest~\cite{Chang1991}, in the last few years much
progress has been made in constructing solid state systems such as
superlattices~\cite{chang_app,Esaki1986,leo2003,Glutsch2004},
quantum wells~\cite{wagner93} and waveguide arrays
\cite{RLGKKZK2003} which enable the controlled observation of RET.
RET has also been examined theoretically for ultracold atoms
trapped in an optical lattice
\cite{GKK1999,Glueck2002,DounasFrazer2007,WMMAKB2005}.

RET-like effects have been observed in a number of experiments to
date. In ref. \cite{Teo2002}, resonant tunneling was observed for
cold atoms trapped by an optical lattice when an applied magnetic
field produced a Zeeman splitting of the energy levels. At certain
values of the applied magnetic field,  the states in the
up-shifting and down-shifting energy levels  were tuned into
resonance with one another. This led to RET drastically altering
the quantum dynamics of the system and producing a modulation of
the magnetization and lifetime of the atoms trapped by the optical
lattice. Resonant tunneling has been observed in a Mott insulator
within an optical lattice, where a finite amount of energy given
by the on-site interaction energy is required to create a
particle-hole excitation \cite{Greiner2002}. Tunneling of the
atoms is therefore suppressed.  If the lattice potential is tilted
by application of a  potential gradient, RET is allowed whenever
the energy difference between neighboring lattice sites due to
the potential gradient matches the on-site interaction energy. The
corresponding nonlinear effect in a Mott insulator allowed
F\"olling {\it et. al.} \cite{Foelling2007} to observe two-atom
RET, and RET in the presence of many-body coherences  was
theoretically analyzed in \cite{Lee2008}. We reported very precise
RET measurements for Bose-Einstein condensates in
\cite{SZLWCMA2007}. The  condensates were loaded into a
one-dimensional optical lattice and subjected to an additional
Stark force, optimally implemented and controlled by accelerating
the lattice.

In this paper, we report additional investigations on RET for a
Bose-Einstein condensate in a one-dimensional optical lattice
applying the high-level control elaborated in our previous work.
The experimental data presented here concentrates on the regime of
parameters for which the Stark force dominates the dynamics of the
condensate.  Our precise control on the experimental parameters
(lattice depth, interaction strength, and the flexibility on the
choice of the initially populated band) enables us to measure the
RET decay of the ground band and the first two excited energy
bands in a wide range of experimental conditions.  Moreover, we
study the impact of atom-atom interactions on the RET process. All
these features extend previous experimental studies on
Landau-Zener tunneling for ultracold atoms in periodic potentials
\cite{BMMWSR1997,Morsch2001,Cristiani2002,Jona2003}. A theoretical
description complements our experimental work.

The  paper is organized as follows. Section 2.1 collects the
necessary theoretical tools to describe our experiments, while
section 2.2 introduces the RET modifications produced by the
atom-atom interactions. Section 3.1 presents our experimental data
in the linear tunneling regime, i.e., in the absence of atom-atom
interactions. The effect of the latter is investigated in section
3.2, before we discuss and summarize our results in section
\ref{concl}.

\section{Theoretical description}
\label{theo}

\subsection{Single-particle RET}
\label{single}

Neglecting for a moment atom-atom interactions in a Bose-Einstein condensate, our system is described by the following Hamiltonian:

\begin{equation}
H = -\frac{\hbar^2}{2M}\frac{d^2}{dx^2} +
V_0 \sin^2\left(\frac{\pi x}{d_L}\right) + Fx \;.
\label{eq:1}
\end{equation}

$V_0$ is the depth of the optical lattice, $d_L$ its spatial period, and $M$ the atomic mass
of Rubidium 87. This Hamiltonian defines the well-known Wannier-Stark problem \cite{N1991,Glueck2002,H2000}.

For small Stark forces $F$, one can picture the evolution of a
momentum eigenstate induced by Eq.~(\ref{eq:1}) as an oscillatory
motion in the ground energy band of the periodic lattice. These
Bloch oscillations with period $T_{\rm Bloch}=h/d_LF$, where $h$
is Planck's constant, were observed for cold and ultracold atoms
in optical lattices \cite{MO2006}.

At stronger forces, a wave packet prepared in the ground-state
band has a significant probability to tunnel at the band edge
(where the band gap $\Delta$ is minimal) to the first excited
band. For a single tunneling event, such a probability is best
estimated by Landau-Zener theory as \cite{H2000}

\begin{equation}
{\rm P_{LZ}} = e^{ -\pi^2 (\Delta/E_{\rm rec})^2/(8F_0) }\;,
\label{eq:2}
\end{equation}

with the recoil energy $E_{\mathrm{rec}}=(\hbar\pi/d_L)^2/2M$ and
$F_0 \equiv Fd_L/E_\mathrm{rec}$. The decay rate -- owing to a
sequence of Landau-Zener tunneling events -- is then obtained by
multiplying ${\rm P_{LZ}}$ with the Bloch frequency
\cite{Glueck2002}

\begin{equation}
\Gamma_{\rm {LZ}} = \nu_{\rm rec} F_0 e^{ -\pi^2 (\Delta/E_{\rm
rec})^2/(8F_0) }\;, \label{eq:3}
\end{equation}

where the recoil frequency is given by $\nu_{\rm rec} =
E_{\rm rec}/\hbar$.

\begin{figure}
\centering
\includegraphics[height=8cm]{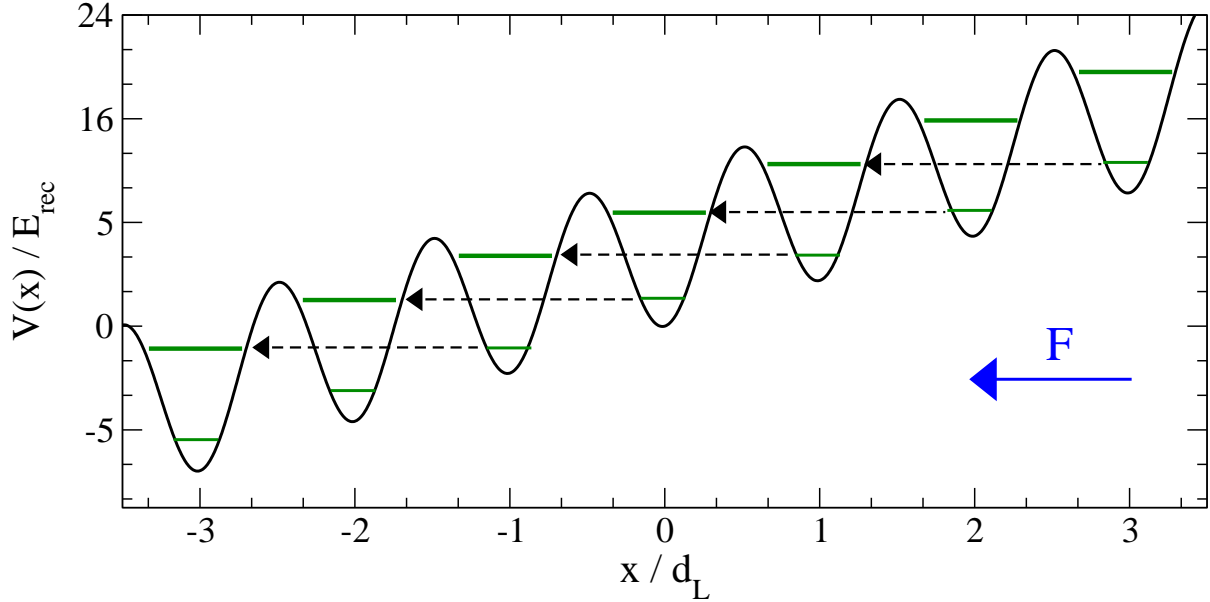}
\caption{Schematic of the RET process between second nearest
neighboring wells, i.e., for $\Delta i=2$. The tunneling of atoms
is resonantly enhanced when the energy difference between lattice
wells matches  the separation between the energy levels in
different potential wells. } \label{fig:1}
\end{figure}

\begin{figure}
\centering
\includegraphics[height=8cm]{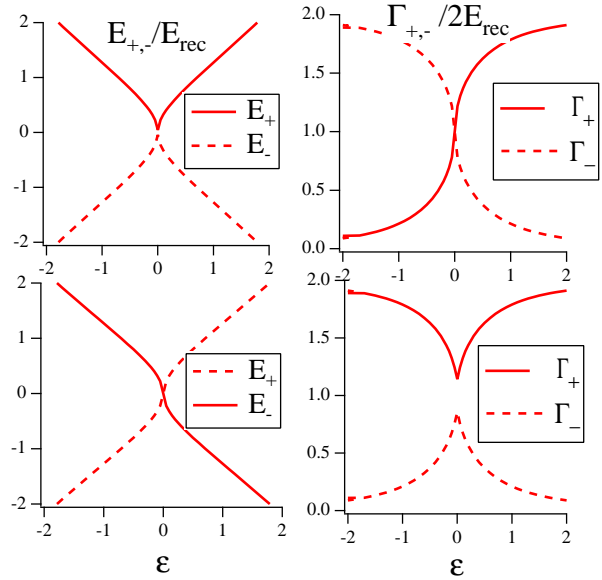}
\caption{Real (left) and imaginary (right) part of the eigenvalues
of Eq. (\ref{eigenvalues}) as a function of $\epsilon$ for
$\gamma=1$, measured in units of $E_\mathrm{rec}$.   A type-I
crossing is found for $v=1.01\,E_{\rm rec}$ (upper plots), and a
type-II crossing is found for  $v=0.99\,E_\mathrm{rec}$ (lower
plots).} \label{fig:2}
\end{figure}

\begin{figure}
\centering
\includegraphics[height=8cm]{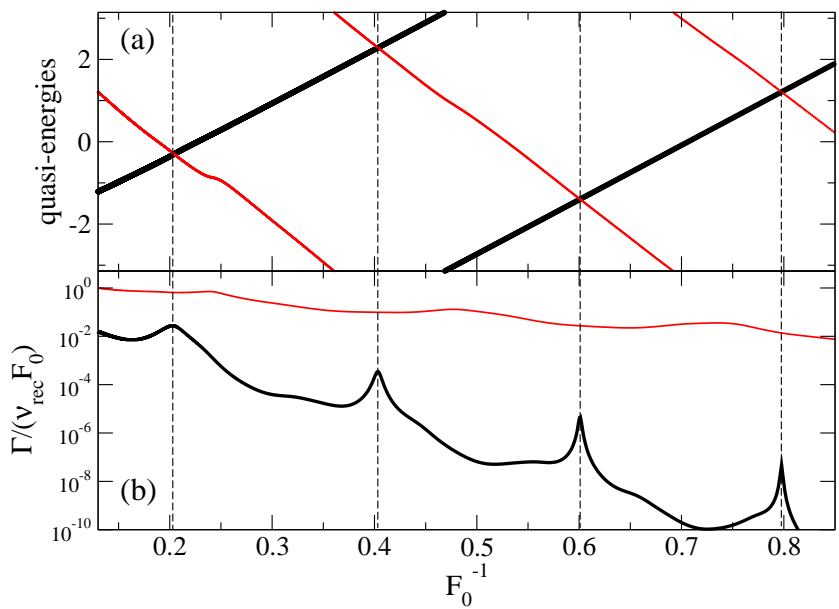}
\caption{In (a) real part of the eigenenergies and in  (b) decay
rates for a lattice depth of $V_0/E_{\rm rec} = 10$ and the
Hamiltonian   from Eq. (\ref{eq:1}). The eigenenergies and the
decay rates are associated with two Wannier-Stark ladders or,
equivalently, with two energy bands: ground state (thick black
lines) and first excited state (thin red lines). The maxima of the
ground-state decay rates correspond to $\Delta i=1, 2$,  $3$, and
$4$.}  \label{fig:1new}
\end{figure}

The actual decay rates can dramatically deviate from Eq.
(\ref{eq:3}) when two Wannier-Stark levels in different potentials
wells are strongly coupled owing to an accidental degeneracy.  By
imposing an energy resonance between the Wannier-Stark levels in
different wells of an optical lattice shifted by the potential of
the external force, one finds that  these degeneracies  occur at
the values $F$ at which $Fd_L \Delta i$ ($\Delta i$ integer) is
close to the mean band gap between two coupled bands of the $F=0$
problem \cite{Glueck2002,Glutsch2004}. The actual peak positions
are slightly shifted with respect to this simplified estimate,
because the Wannier-Stark levels in the potential wells are only
approximately defined by the averaged band gap of the $F=0$
problem, a consequence of field-induced level shifts
\cite{Glueck2002}. The RET process based on the $n=1$ and $n=2$
levels of the Wannier-Stark ladder is sketched in Fig. 1.

The modification of the level decay rate by the presence of a  degeneracy may be described  by a simple model of
 a two-level Hamiltonian with energy splitting $2 \epsilon$ and one decaying level \cite{Avron82,Keck03}
\begin{eqnarray}
H= \left[
\begin{array}{cc}
 \epsilon -i \gamma & v  \\
 v & - \epsilon
\end{array}
\right]\,.
\label{hamiltonian}
\end{eqnarray}

In this approach it is assumed that the upper bare state decays
with rate $\gamma$, while the decay is negligible for the other
one. The two states are coupled with strength $v$. The eigenvalues
of the non-hermitian Hamiltonian of Eq. \ref{hamiltonian} are
given by

\begin{equation}
{\cal E}_{\pm}=-i\gamma \pm\sqrt{\left(\epsilon-i\gamma\right)^2 + v^2} = E_{\pm}-i\Gamma_{\pm}/2.
\label{eigenvalues}
\end{equation}

Real and imaginary part of the eigenvalues are different for
$\epsilon \ne 0$, but  crossings or anticrossings of the real and
imaginary part are found at the critical value $\epsilon = 0$
where two different scenarios take place.  For $|v| \ge\gamma$, at
$\epsilon = 0$  the imaginary parts of the eigenvalues coincide,
$\Gamma_+ = \Gamma_- = 2\gamma$, while the real parts differ. In
this case,  denoted as type-I crossing, the imaginary parts of the
eigenvalues cross while the real parts anti-cross, as shown in the
upper plots of Fig. 2. For $|v| \le \gamma$, at $\epsilon = 0$ the
real parts of the eigenvalues coincide, $E_+= E_-= 0$, while the
imaginary parts differ. In this case, denoted as type-II crossing,
the eigenvalues anticross while the real parts cross, as shown in
the lower plots of Fig. 2. Type-II  crossing corresponds to the
RET phenomenon: if the lower state is energetically close or equal
to the decaying upper level, the decay rate of the lower state
increases significantly. In addition the upper state experiences a
resonantly stabilized tunneling (RST) with a decrease of its decay rate.

For non-interacting atoms described by Eq. (\ref{eq:1}), we can
easily diagonalize an opened version of our Hamiltonian
\cite{GKK1999,Glueck2002,WSM2006,SW2007,WGWK2007} to obtain the
true resonance eigenstates and eigenenergies of our decaying
system. Fig. \ref{fig:1new}(a) shows the crossing and
anticrossings for the real parts of the eigenenergies  associated
with a  configuration investigated experimentally as a function of
the experimental control parameter, the Stark force. It may be
noticed that type-II crossings are typically encountered for our problem
of decay from lower bands. The associated Wannier-Stark states decay with
rates are shown in Fig. \ref{fig:1new}(b) as a function of the
dimensionless parameter $F_0$. The strong
modulations on top of the global exponential decrease arise from
RET processes originated by the type-II crossings.

\subsection{Interacting Bose-Einstein condensate dynamics}
\label{int}

\begin{figure}
\centering
\includegraphics[height=5cm]{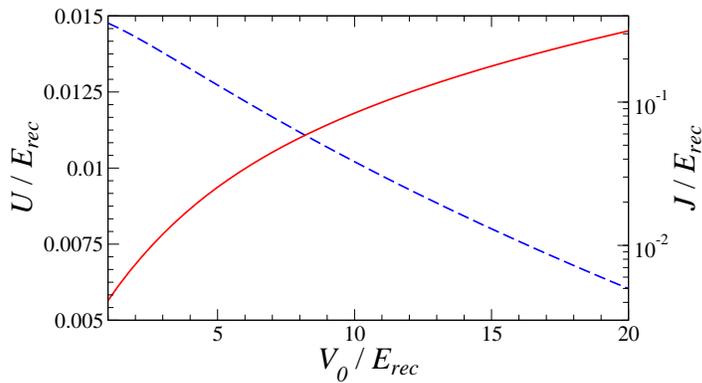}
\caption{Hopping parameter $J$ (dashed line) and on-site interaction constant $U$ (solid line)
         of a 1D Bose-Hubbard model as a function of the depth of the optical lattice.
         $U$ is computed for Rubidium 87 and for typical experimental parameters, i.e., lattice spacing
         $d_L=620\rm\,nm$, and  radial confinement  frequency $\omega_r/2\pi=250\rm\,Hz$,
         using the projection to a quasi-1D situation of \cite{BMO2003}.
}
\label{fig:3}.
\end{figure}

In this section we discuss the effect of atom-atom interactions in
the Bose-Einstein condensate and how to effectively model them for
a quantitative description of the experiment. We focus on a
parameter regime where the Stark force essentially dominates the
dynamics of the condensate. Here the quantum tunneling between the
energy bands is significant and most easily detected
experimentally. The critical field values for which such
excitations are relevant can be estimated by comparing, for
instance, the potential energy difference between neighboring
wells, $F d_{\rm L}$, with the coupling parameters of the
many-body Bose-Hubbard model, i.e., the hopping constant $J$ and
interaction constant $U$\cite{MO2006}. These parameters are
plotted in Fig. \ref{fig:3} for typical experimental parameters
as a function of the lattice depth $V_0$.

Our theoretical and experimental analysis will exclude the regime
of  $F_0 \le J/E_{\rm rec}$ where a quantum chaotic system is
realized\cite{BK2003,TGZ2003,TMW2007,TMW2007a}.  The origin of
ÔÔquantum chaos,ÕÕ i.e., of the strongly force-dependent,
nonperturbative mixing of energy levels can be understood as a
consequence of the interaction-induced lifting of the degeneracy
of the multiparticle Wannier-Stark levels in the crossover regime
from Bloch to Wannier spectra, making nearby levels strongly
interact, for comparable magnitudes of hopping matrix elements and
Stark shifts.

For the regime of $F_0 \gg J/E_{\rm rec}$ studied here, the effect
of weak interactions is just a perturbative shifting and a small
splitting of many-body energy levels \cite{WSM2006,TMW2007a}. As a
consequence, we can use a global mean-field description based on
the Gross-Pitaevskii equation \cite{BEC_books} to simulate the
temporal evolution of a Bose-Einstein condensate wave function
$\psi(\vec{r},t)$ subjected to a realistic potential

\begin{eqnarray}
\fl \ii \hbar \frac{\partial }{\partial t}\psi (\vec{r},t) = \nonumber
\\ \hspace{-1.5cm}
\left[-\frac{\hbar^2}{2M}\nabla ^2 + \frac{1}{2}M \left( \omega_x^2 x^2 +
\omega_{\rm r}^2 \rho^2 \right) + V_0 \sin^2\left(\frac{\pi x}{d_L}\right)
+ F x + g \left| \psi(\vec{r},t) \right|^2
\right] \psi(\vec{r},t).
\label{eq:4}
\end{eqnarray}

The frequencies $\omega_x$ and $\omega_{\rm r}$ characterize the
longitudinal and transverse harmonic confinement (with cylindrical
symmetry of the optical dipole trap: $\rho = \sqrt{y^2 + z^2}$, cf. section \ref{exp}).
The atom-atom interactions
are modeled by the nonlinear term in Eq. (\ref{eq:4}),with the
nonlinear coupling constant  given by $g=4\pi \hbar^2 a_s/M$,
where $a_s$ is the $s$-wave scattering length \cite{BEC_books}.
Later, we will use the dimensionless nonlinearity parameter
$C=gn_0/(8E_{\rm rec})$\cite{Cristiani2002,MO2006}, which is
computed from the peak density $n_0$ of the initial state of the
condensate, to describe the experimentally relevant nonlinear
couplings $C \approx 0.01 \ldots 0.06$. In the Thomas-Fermi regime
of the condensate~\cite{BEC_books}, for given $\omega_x$ and
$\omega_r$ the density $n$, and therefore $C$, is proportional to
$N^{2/5}$ where $N$ is the number of atoms in the condensate.

The Gross-Pitaevskii equation (\ref{eq:4}) is numerically integrated using finite difference
propagation, amended by predictor-corrector loops to reliably
evolve the nonlinear interaction term \cite{M1998}.
To avoid any spurious effects owing to the fast spreading of the wave functions,
we use a large numerical basis, especially in the longitudinal direction. In this way, we fully cover
the 3D expansion of the entire wave packet, including its tunneled tail,
without the use of non-Hermitian potentials. The initial state propagated by Eq. (\ref{eq:4})
is the relaxed condensate wave function, adiabatically loaded into the
confining potential given by the harmonic trap and the optical lattice (at $F=0$).

In oder to have access to the decay rates in the experiment, one
needs to measure the temporal evolution of the probability of the
condensate to remain in the energy band, in which it has been
prepared initially. As proposed in \cite{WMMAKB2005}, such a
survival probability is best measured in momentum space, since,
experimentally, the most easily measurable quantity is the
momentum distribution of the condensate obtained from a free
expansion after the evolution inside the lattice. From the
momentum distributions we determine the survival probability by
projection of the evolved state $\psi(\vec{p},t)$ onto the support
of the initial state:

\begin{equation}
P_{\rm sur}(t) \equiv \int_{-p_{\rm c}}^{p_{\rm c}} dp_x
\left( \int dp_ydp_z |\psi(\vec{p},t)|^2 \right)\;.
\label{eq:5}
\end{equation}

A good choice is $p_{\rm c} \geq  3p_{\mb R}$ since typically
three momentum peaks are initially significantly populated when
loading the condensate adiabatically into the periodic lattice,
and they correspond to $-2p_{\mb R},0,2p_{\mb R}$
\cite{Morsch2001,Cristiani2002,MO2006}. For $g=0$, the individual
tunneling events occurring when the condensate crossed the band
edge are independent, and hence $P_{\rm sur}(t)$ has a purely
exponential form (apart from the $t \to 0$ limit
\cite{WBFMMNSR1997}). When the nonlinear interaction term is
present, the density decays with time too. As a consequence, the
rates $\Gamma$ are at best defined locally in time, and in the
presence of RET even a sharp non-exponential decay is possible, as
discussed in \cite{Carr2005,SW2007}. Nevertheless, for the short
evolution times and the weak nonlinear coupling strengths $C$ that
are experimentally accessible, the decay of the condensate can be
well fitted by an exponential law \cite{WCMMA2007,SZLWCMA2007}

\begin{equation}
P_{\rm sur}(t) = P_{\rm sur}(t=0) \exp \left(- \Gamma_n t \right) = \exp \left(- \Gamma_n t\right) \;,
\label{eq:surexp}
\end{equation}
with rates $\Gamma_n$ for the band $n=1$ (ground band), 2 (first excited band), 3 (second excited band), in which the atoms are initially prepared.

Before we discuss our experimental setup and present our data on
linear and nonlinear tunneling, we come back to the RET peaks
discussed above, c.f., Fig. \ref{fig:1}. These peaks, which are
predicted to occur for the single-particle motion studied in
section \ref{single}, will be affected by the nonlinear
interaction term of Eq. (\ref{eq:4}). A shift of the RET peaks in
energy or in the position of the Stark force, as predicted in
\cite{WSM2006} for much larger parameters $C$, is negligible for
our nonlinearities $C < 0.06$, for which such a shift would
correspond to the extremely small amount of $\Delta F_0 < 5 \times
10^{-4}$ \cite{WSM2006}. The RET peaks, however, originate from an
exact matching of energy levels in neighboring potential wells,
and hence they are very sensitive to slight perturbations. We may
estimate the necessary perturbation by the nonlinear term in Eq.
(\ref{eq:4}) by comparing the width of the RET peaks of a band $n$
(which essentially is determined by the decay width  $\Gamma_{\rm n+1}$ of the band into
which the atoms tunnel) with the energy scale of the
nonlinearity. In the experiment we can easily reach nonlinearities corresponding to
this order-of-magnitude argument, and the consequences will be
discussed in section \ref{nonl} below.

\section{Experimental results}
\label{exp}

The starting point of the measurements presented in
this article is the creation of a Bose-Einstein condensate (BEC)
of $^{87}$Rb atoms. This is realized starting from a cloud of
atoms trapped in a 3D magneto-optical trap (MOT) and then loaded
in a pure-magnetic time-orbital potential (TOP) trap after a
molasses stage for sub-Doppler cooling. In order to achieve
condensation, evaporative cooling is performed first in the TOP
trap and then in an all-optical dipolar trap, where the atoms are
transferred once they have a temperature of few $\mu K$. A BEC of
up to $5\times 10^4$ atoms then forms in the optical trap. The
dipolar trap is realized with two off-resonant Gaussian laser
beams focused to waists of $50\,\mathrm{\mu m}$, having a
wavelength $\lambda=1030\,\mathrm{nm}$, and mutually detuned by
$\sim220\,\mathrm{MHz}$ in order to avoid interference. The aspect
ratio of the trap can be varied through the power of the laser
beams, which is up to $1\,\mathrm{W}$ each and actively controlled
by a feed-back loop. This feedback loop permits us to decrease the
intensity noise on the beams and to improve reproducibility during
the data collection.

After the creation of the condensate, the trap frequencies are
adiabatically varied in order to confine the BECs in a
cigar-shaped trap, with a longitudinal frequency of $\sim
20\,\mathrm{Hz}$ and radial frequency in the range
$80-250\,\mathrm{Hz}$. The BECs are then loaded into a
one-dimensional optical lattice oriented along the weak direction
of the dipolar trap. The lattice is created by optical
interference of two linearly polarized Gaussian laser beams
($\lambda = 852\,\mathrm{nm}$) focused to a waist of
$120\,\mathrm{\mu m}$ and intersecting with an angle $\theta$. The
lattice spacing is then  $d_{L}=\lambda/(2 \sin (\theta/2) )$. The lattice depth  $V_0$
 is controlled through the laser
intensity, and will be expressed in units of the recoil energy
$E_{\rm rec}$. The measurements presented
in the article were taken for different values of the lattice
depth and of the lattice spacing: $V_{0}/E_{rec}=6, 4, 9, 16$ with
$d_{L}=0.426\mu m$, and $V_{0}/E_{rec}=2.5, 10, 12, 14$ with
$d_{L}=0.620\,\mathrm{\mu m}$.
Each lattice beam passes through an acousto-optic modulator (AOM)
in order to control its power and hence the lattice depth.
Moreover, by varying the radio-frequency driving one of the two
AOMs, it is possible to create a detuning $\Delta \nu$ between the
two lattice beams. This causes a displacement in time of the
lattice in the laboratory frame. Within this frame, it is possible
to make the lattice move at a velocity $v=d_{L} \Delta\nu$, or to
accelerate it with an acceleration $a=d_{L}(d\Delta \nu/dt)$.

The lattice is usually loaded in $1\,\mathrm{ms}$ to avoid
excitations to higher bands, and the atoms occupy the fundamental
band if they have zero group velocity in the lattice rest frame
during the loading phase. However, if the lattice is loaded with a
constant velocity, the atoms can occupy one of the excited bands
if the energy and quasi-momentum are conserved~\cite{Peik1996}.
Furthermore, when the lattice is accelerated, the atoms are
subjected to a force $F=ma$ in the rest frame of the lattice: this
corresponds to the experimental realization of the Hamiltonian
(1). The applied force $F$ is chosen in order to minimize the
growth of dynamical instabilities, as explored in
\cite{Cristiani2004}.

In order to measure the tunneling rate $\Gamma_n$ for BECs
initially loaded into the $n$-th band of the optical lattice
(ground state: $n=1$, first excited state: $n=2$, etc.), the
lattice is accelerated with acceleration $a$ for an integer number
of Bloch oscillation cycles. During this acceleration time, atoms
are most likely to tunnel to upper bands when the condensate
quasi-momentum is close to the edge of the Brillouin zone. Atoms
that do not tunnel to a higher band and are, therefore, "dragged
along" by the accelerated lattice acquire a larger final velocity
than those that have undergone tunneling. They are spatially
separated from the latter by releasing the BEC from the dipole
trap and lattice at the end of the acceleration period and
allowing it to expand and to fall under gravity for
$5-20\,\mathrm{ms}$. After the time-of-flight, the atoms are
detected by absorptive imaging on a CCD camera using a resonant
flash.

>From the dragged fraction $N_{drag}/N_{tot}$, we then determine
the tunneling rate $\Gamma_n$ by imposing the asymptotic decay law
\begin{equation}
N_{drag}(t)=N_{tot}\exp{(-\Gamma_n t)}
\end{equation}
where the
subscript $n$ indicates the dependence of the tunneling rate on
the local energy level $n$ in which the atoms are initially
prepared. Our measurement of $\Gamma_n$ based on the dragged
fraction relies on the fact that for the lattice depths used in
our experiments the number of bound states in the wells was small
(2-4, depending on the lattice depth), so after the first
tunneling event, the probability for tunneling to the next bound
state or the continuum was close to unity. This explains why we
observe type-II crossings, corresponding to $\gamma > |v|$ in the
model discussed in section \ref{single}.

The way in which we measure the tunneling rate also determines the
achievable resolution of our method. This is given by the minimum
number of atoms that we can distinguish reliably from the
background noise in our CCD images, which varies between $500$ and
$1000$ atoms, depending on the width of the observed region. With
our condensate number, and taking into account the minimum
acceleration time limited by the need to spatially separate the
two fractions after time-of-flight and the maximum acceleration
time limited by the field of view of the CCD camera, this results
in a maximum $\Gamma_n/\nu_{\mathrm rec}$ of $\approx 1$ and a
minimum of $\approx 1\times 10^{-2}$.

\subsection{The linear regime}
\label{linear}

Although the finite and positive scattering length of the 87-Rb
atoms in our BECs means that the linear
Hamiltonian of Eq.(\ref{eq:1}) is never exactly realized in our
experiments, we can approximate a non-interacting BEC by keeping
the condensate density low. In that case, the interaction energy
can be made much smaller than all the other energy scales of the
system (recoil energy, band width, gap width) and hence negligible
for our purposes. A low density can be achieved by using a weak
trap with small trap frequencies and/or a small atom number in the
BEC. Alternatively, one can also allow the BEC to expand freely
for a short time (typically less than a millisecond, to avoid
excessive dropping under gravity) before performing the lattice
acceleration.

\begin{figure}
\centering
\includegraphics[height=16cm]{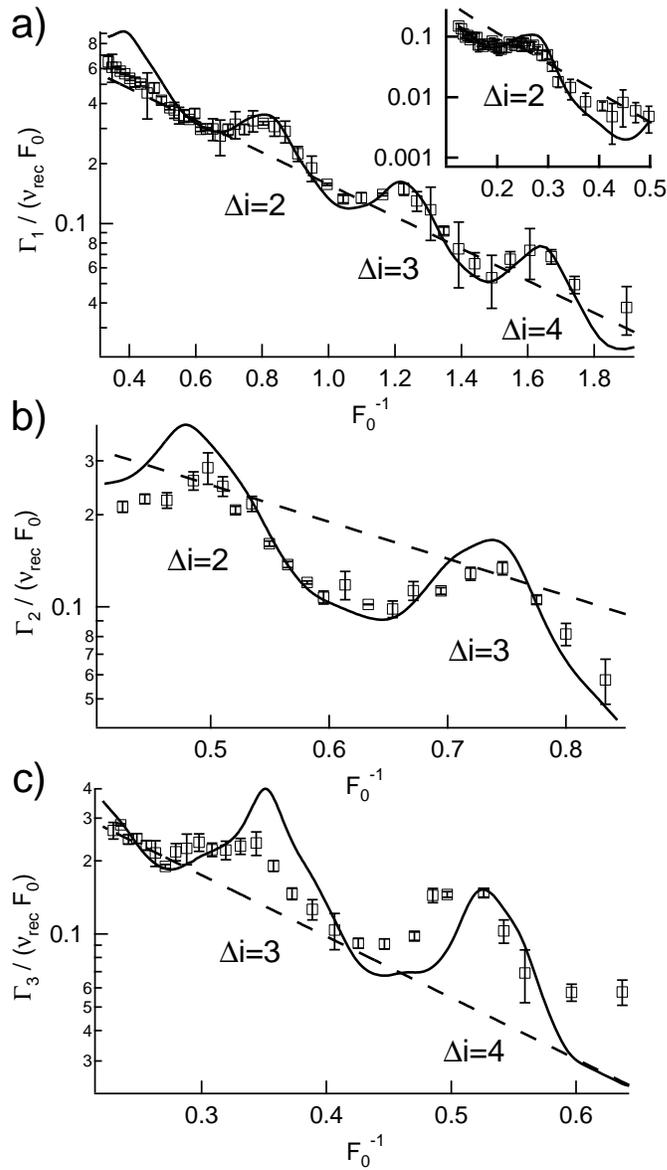}
\caption{Resonant tunneling in the linear regime. Shown here are
the tunneling rates from the three lowest energy bands of the
lattice as a function of the normalized inverse force $F_0^{-1}$
for lattice depths (a) $V_0/E_\mathrm{rec}=2.45$ (insert:
$V_0/E_\mathrm{rec}=6$), (b) $V_0/E_\mathrm{rec}=10$ and (c)
$V_0/E_\mathrm{rec}= 23$. Note the different scales on the
horizontal axes.} \label{lin}.
\end{figure}

Fig.~\ref{lin} shows the results of experiments with low-density
condensates for which the nonlinearity parameter $C$ was less than
$\approx 1\times 10^{-2}$, which in this work we define to be the
limit of the linear regime. In each plot, the tunneling rate
$\Gamma_{n}$ out of the $n$-th band (in our experiments we were
able to study the cases $n=1,2$ and $3$) is shown as a function of
$F_0^{-1}$. Superimposed on the overall exponential decay of
$\Gamma_n/F_0$ with $F_0^{-1}$, one clearly sees the resonant
tunneling peaks corresponding to the various resonances $\Delta
i=1,2,3,4$. Which of the resonances were visible in any given
experiment depended on the choice of lattice parameters and the
finite experimental resolution. The limit $n=3$ for the highest
band we could explore was given by the maximum lattice depth
achievable.

\begin{figure}
\centering
\includegraphics[height=14cm]{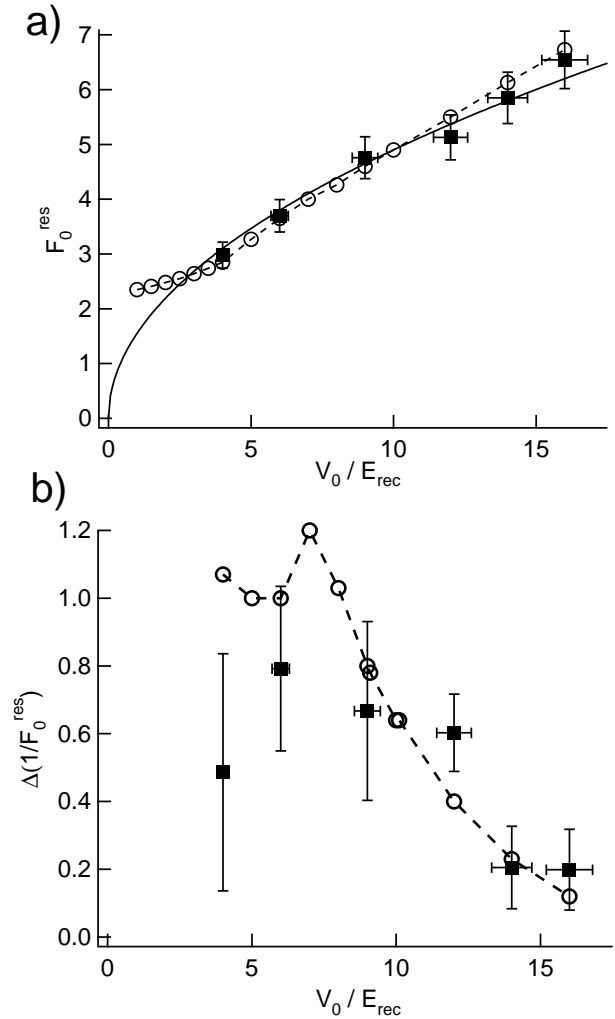}
\caption{Positions (a) and widths (b) of the tunneling resonances
with $\Delta i = 1$ in the lowest energy band as a function of the lattice depth. In
(a), the dashed line is the theoretical prediction based on the
harmonic oscillator approximation, modified as described in the
main text. In (a) and (b), the open symbols connected by the
dashed line are the results of a numerical simulation.}
\label{widths}.
\end{figure}

The inset in Fig.~\ref{lin} (a) shows the tunneling resonances in
the lowest energy band for a different value of the lattice depth
$V_0$. One clearly sees that the positions of the resonances are
shifted according to the variation in the energy levels.
Fig.~\ref{widths} (a) shows the positions {\bf $F^{\rm res}_0$} of
the $\Delta i=1$ resonances as a function of the lattice depth.
For deep enough lattices, these positions agree perfectly with the
results of a numerical simulation (see Fig. 6 (a)) and can also be
approximately calculated by making a harmonic approximation in the
lattice wells, which predicts a separation of the two lowest
energy levels ($n=1$ and $n=2$) of

\begin{equation}
\Delta E_{2-1} = 2E_\mathrm{rec}\sqrt{\frac{V_0}{E_\mathrm{rec}}}.
\end{equation}

The resonance condition $\Delta E_{2-1}=F^{\rm res}  d_L \Delta i$
can then be used to calculate the resonance position $F^{\rm
res}$. Our experimental results of Fig.~\ref{lin} (a) are well
fitted by this formula if the factor $2$ in the expression for
$\Delta E_{2-1}$ is replaced by $\approx 1.5$. This discrepancy
with the theoretical prediction is to be expected since the
anharmonicity of the potential wells reduces the actual energy
separation of the levels compared to the harmonic case.
While we were not able to measure the tunneling resonances in two
\textit{different} bands for the \textit{same} lattice depth, we
could measure the resonances in one single band and compare our
results with the theoretically predicted resonances in an adjacent
band~\cite{SZLWCMA2007}. This allowed us to confirm that in our
experiments a resonance peak in one band always coincided with an
anti-peak or trough in the adjacent band, which agrees with our
interpretation in terms of a type-II crossing (see section 2.1).

We also studied the dependence of the widths of the tunneling
resonances on the lattice depth. Physically, this width is
determined by the width of the state to which the atoms tunnel and
hence should decrease with increasing lattice depth. For instance,
for tunneling from the ground state band $n=1$, the resonance
width should reflect the width of the first excited band $n=2$.
Figure~\ref{widths} shows the results of our measurements. For
large lattice depths, the resonance width decreases as expected,
whereas for shallow lattices the behaviour is more complicated.
This is also reflected in the numerical simulations.

\subsection{The nonlinear regime}
\label{nonl}

In order to enter the regime for which $C\gtrsim 1\times 10^{-2}$,
we carry out the acceleration experiments in radially tighter
traps (radial frequency $\gtrsim 100\,\mathrm{Hz}$) and hence at
larger condensate densities. Fig.~7 shows the $\Delta i=2$ and
$\Delta i=3$ resonance peaks of the ground-state band ($n=1$) for
increasing values of $C$, starting from the linear case and going
up to $C\approx3\times 10^{-2}$. As the nonlinearity increases,
two effects occur. First, the overall (off-resonant) level of
$\Gamma_1$ increases linearly with $C$. This is in agreement with
our earlier experiments on nonlinear Landau-Zener
tunneling~\cite{Morsch2001,Jona2003} and can be explained
describing the condensate evolution within  a
nonlinearity-dependent effective potential
$V_{\mathrm{eff}}=V_0/(1+4C)$~\cite{choiniu1999}. Second, with
increasing nonlinearity, the contrast of the RET peak is decreased
and the peak eventually vanishes, as is also evident from the
different on-resonance and off-resonance dependence of the
tunneling rate as a function of the atom number $N$ (and hence the
nonlinearity), as seen in Fig. 7 (b). This is in agreement with
the theoretical discussion of section 2.

As mentioned in section 2.2, the critical value of $C$ for which
the nonlinearity significantly affects the resonance peak should
be given by the width of the resonance peak itself. For the parameters of Fig.~5 (a)
and 7 (a) and the RET peak with $\Delta i = 2$, the typical width $\Gamma_2$ of the decaying state to
which the atoms tunnel is of the order of $0.2 \ldots 0.5$, expressed in
units of $E_\mathrm{rec}$. Since $C$ reflects the nonlinearity
expressed in units of $8\times E_\mathrm{rec}$, this means that we
expect to see substantial deviations from the linear behaviour
when $C\gtrsim 0.025 \ldots 0.06$. Experimentally, we confirm that this
threshold is a good estimate for the onset of the destruction of the
RET peak, which is observed to occur around $C=0.02$ in Fig. 7 (a).

\begin{figure}
\centering
\includegraphics[height=18cm]{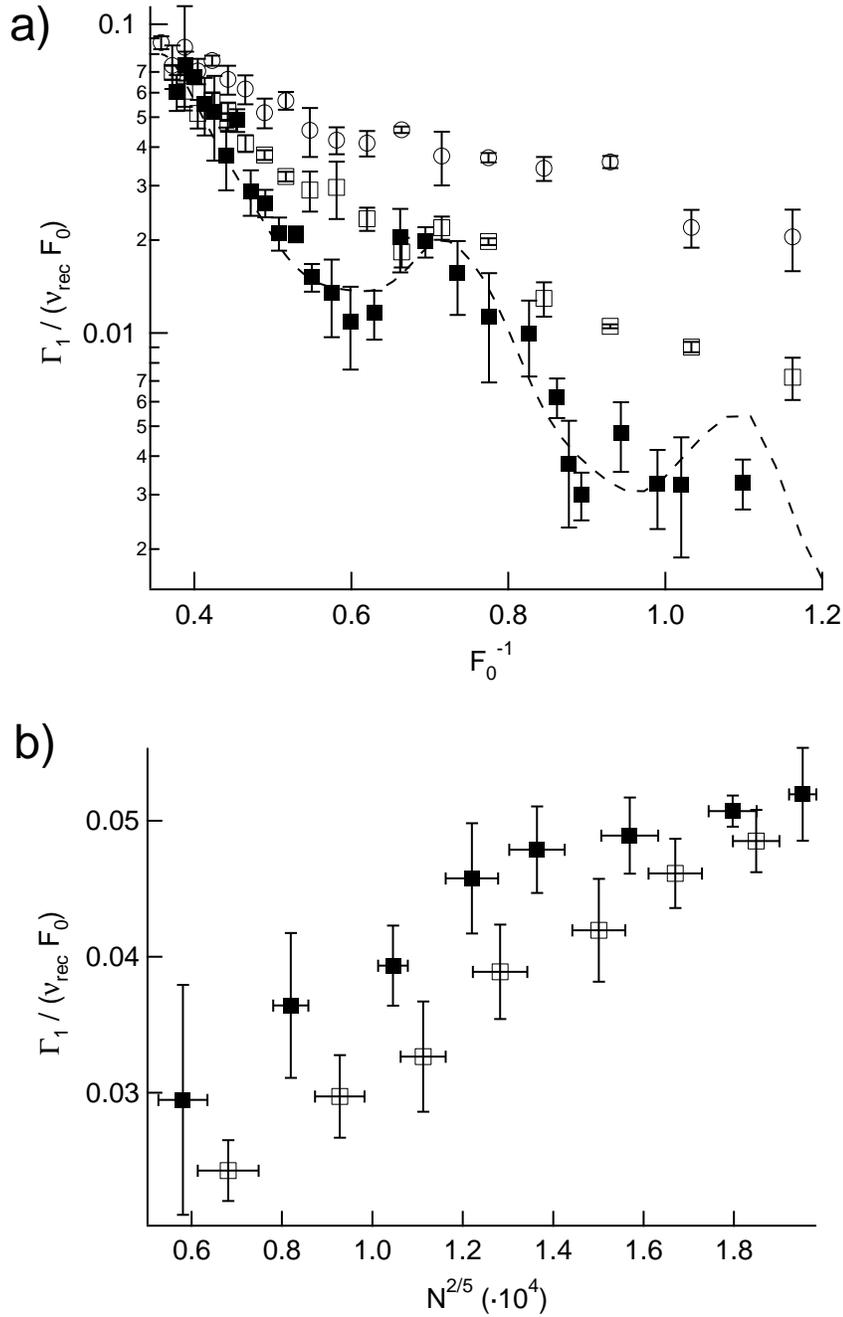}
\caption{Resonant tunneling in the nonlinear regime. (a) The
tunneling rates for $\Delta i = 2$ from the lowest energy band of the optical lattice
as a function of the normalized inverse force $F_0^{-1}$ for a
lattice depth  $V_0/E_\mathrm{rec}=3.5$  and different values of
$C\approx 0.01, 0.022,0.033$ from bottom to top. The dashed line
is the theoretical prediction in the linear regime. As the
nonlinearity increases, the overall tunneling rate increases and
the resonance peak becomes less pronounced. (b) Dependence on the
condensate atom number $N$ of the tunneling rate at the position
of the peak $F_0^{-1} = 0.71$ (solid symbols) and of the through
$F_0^{-1} =0.60$ (open symbols) for $V_0/E_\mathrm{rec}=3.0$. }
\label{nonlin}.
\end{figure}

\section{Conclusions and outlook}
\label{concl}

In this paper, we have studied the resonantly enhanced
tunneling of BECs in optical lattices both theoretically and
experimentally. Our results show that ultracold atoms in periodic
potentials are well suited to simulating and exploring basic
quantum mechanical processes which are also the subject of active
investigations in the solid state physics community, such as Bloch
oscillations~\cite{Waschke93,Feldmann1992,Pertsch1999,Morandotti1999}
and Zener tunneling~\cite{Ghulinyan2005,Trompeter2006}. Compared
to solid-state experiments, our approach offers the advantage of a
large flexibility in the experimental parameters and the
possibility to add a nonlinearity in a controlled way.

The experimental setup presented in this paper also opens up the
possibility to explore different regimes, such as the strongly
interacting regime for $J \simeq U \gtrsim F_0$
\cite{TMW2007,TMW2007a}. Another interesting aspect to be studied
in the nonlinear regime is the limit in which the fraction of
atoms undergoing tunneling is either very large (i.e., very few
atoms remain in the initial band) or very small. In both limits,
deviations from the the Gross-Pitaevskii equation, which
presupposes a mean-field approximation for all the bands involved,
are expected~\cite{Shchesnovich_07}.

\ack This work was supported by the European STREP Project OLAQUI,
a MIUR PRIN-2005 Project, and the Sezione di Pisa dell'INFN. S.W.
acknowledges support from the Alexander von Humboldt Foundation
(Feodor-Lynen Program 2004-2006) and within the framework of the
Excellence Initiative by the German Research Foundation (DFG)
through the Heidelberg Graduate School of Fundamental Physics
(grant number GSC 129/1), as well as a travel grant from CNISM
Unit\`{a} di Pisa. The authors would like to thank Matteo
Cristiani for assistance, and Lincoln Carr, Andrey Kolovsky, Hans-J\"urgen Korsch
and Peter Schlagheck for enlightening discussions.

\end{document}